\documentstyle[seceqn,equation]{espart}
\font\st=eufm10 scaled 1200
\font\bi=cmbxti10 scaled 1200
\font\bfs=cmmib10 scaled 1200
\font\bfd=cmbsy10 scaled 1200
\def\bs{\hbox{\bfs\symbol{27}}}
\def\bd{\hbox{\bfd\symbol{1}}}
\def\bt{\hbox{\bfd\symbol{2}}}
\begin{document}
\begin{frontmatter}
\title{\hfill Groups and nonlinear dynamical systems\hfill}
\subtitle{\hfill\bf Chaotic dynamics on the {\bi SU(2)}\bt {\bi SU(2)} group\hfill}
\author{\hfill{\bf KRZYSZTOF KOWALSKI} and {\bf JAKUB REMBIELI\'NSKI}\hfill}
\address{Department of Theoretical Physics, University
of \L\'od\'z, ul. Pomorska 149/153,\\ 90-236 \L\'od\'z, Poland}
\begin{abstract}
In our previous paper: K. Kowalski and J. Rembieli\'nski, Groups
and nonlinear dynamical systems.  Dynamics on the $SU(2)$ group,
{\em Physica D\/} {\bf 99}, 237 (1996), we introduced an abstract Newton-like equation
on a general Lie algebra such that submanifolds fixed by the
second-order Casimir operator are attracting set.  The corresponding
group parameters satisfy the nonlinear dynamical system having an attractor
coinciding with the submanifold.  In this work we discuss the case with
the $SU(2)\times SU(2)$ group.  The resulting second-order system in
${\bf R}^6$ is demonstrated to exhibit chaotic behaviour.
\end{abstract}
\end{frontmatter}

\vspace{1.2cm}
Keywords:\qquad \parbox[t]{13cm}{dynamical systems, ordinary
differential systems, Lie groups,\\ deterministic chaos, symmetric
attractors}

\vspace{.8cm}
\newpage
\setcounter{section}{1}
\centerline{\bf 1.\ INTRODUCTION}

\vspace{\baselineskip}
One of the most important problems of the theory of dynamical
systems relates to finding and clasifying attracting limit sets or
simply attractors.  Another important problem is how to study the
system on its invariant manifold.  In our previous work [1] we
introduced a method for the construction of $G$-invariant nonlinear
dynamical systems having an attracting set coinciding with a
submanifold generated by the second-order Casimir operator referring to
a given Lie group $G$.  The example of the $SU(2)$ group
discussed therein led to the oscillatory dynamics on the orbit i.e.\
the sphere $S_2$.  As suggested in [1] such regular dynamics need
not be the case for groups with higher-dimensional submanifolds.  In this
paper we examine the case with the $SU(2)\times SU(2)$ group.
Namely, following the general scheme described in [1] we introduce
the nonlinear second-order system in ${\bf R}^6$ satisfied by the
group parameters.  We then show that the system exhibits chaotic
behaviour on the sphere $S_5$ which is a submanifold fixed by the
second-order Casimir operator corresponding to the $SU(2)\times SU(2)$ group.

\vspace{1.5\baselineskip}
\setcounter{section}{2}
\centerline{\bf 2.\ THE NEWTON-LIKE EQUATION}

\vspace{\baselineskip}
In this section we recall the abstract Newton-like
equation on a general Lie algebra such that submanifolds fixed by
the second-order Casimir operator are attracting set [1].  This equation
generates the nonlinear dynamical system satisfied by the group parameters,
having an attractor coinciding with the submanifold.  Consider the following
second-order  differential equation on a Lie algebra $\hbox{\st\symbol{103}}$:
\begin{equation}
\mu \ddot X + \nu\dot X +  \rho X + \sigma Y =
e^{iX}Ye^{-iX},\qquad X(0)=X_0,\quad \dot X(0)=\dot X_0,
\end{equation}
where $X(t)$: ${\bf R}\to \hbox{\st\symbol{103}}$ is a
curve in $\hbox{\st\symbol{103}}$,
$Y\in\hbox{\st\symbol{103}}$ is a fixed element, $\mu,
\,\nu,\,\rho,\,\sigma\in{\bf R}$ and dot designates
differentiation with respect to time.

Let us assume that $\mu\ne0$.  Clearly, we can set without loose of
generality $\mu >0$.  On demanding that (2.1) admits the solution on
the submanifold fixed by the second-order Casimir operator of the form
\begin{equation}
{\rm Tr}X^2={\rm const}\ne 0,
\end{equation}
rescaling $t\to \sqrt{\mu}\,t$ and setting $\frac{\nu}{\sqrt{\mu}}=\beta$
we arrive at the following Newton-like equation:
\begin{equation}
\ddot X + \beta \dot X + \frac{{\rm Tr}{\dot X}^2}{{\rm
Tr}X^2 }X = e^{iX}Ye^{-iX} - Y,\qquad X(0)=X_0,\quad \dot
X(0)=\dot X_0.
\end{equation}
Using (2.3) we find
\begin{equation}
{\rm Tr}X^2 = 2{\rm Tr}X_0\dot
X_0\hbox{$\frac{1}{\beta}$}(1-e^{-\beta t}) + {\rm
Tr}X_0^2.
\end{equation}
That is if $\beta > 0$ then the solution
to (2.1) approaches the submanifold
\begin{equation}
{\rm Tr}X^2 = \hbox{$\frac{2}{\beta}$}{\rm Tr}X_0\dot
X_0 + {\rm Tr}X_0^2.
\end{equation}
It can be easily checked that for $\beta \le0$ there is no solution
on the submanifold.  The only exception are the initial data such that
\begin{equation}
{\rm Tr}X_0\dot X_0 = 0,
\end{equation}
The more detailed analysis of eq.\ (2.3) and (2.4) is provided in
[1].   Notice that in the general case of an $n$-dimensional compact Lie
algebra the manifold defined by (2.2) is simply the sphere $S_{n-1}$.

\vspace{1.5\baselineskip}
\setcounter{section}{3}
\centerline{\bf 3.\ DYNAMICS ON THE {\bi SU}(2)\bt{\bi SU}(2) GROUP}

\vspace{\baselineskip}
Our aim now is to discuss the nonlinear dynamical system implied by
the abstract Newton-like equation in the case of the
$SU(2)\times SU(2)$ group.  We first observe that due to the product
structure of the considered group, the general elements of the Lie
algebra $X(t)$ and $Y$ can be written in the form
$$\displaylines{
\hspace{1em}X(t) = X_+(t)+X_-(t),\hfill\cr
\hspace{1em}Y = Y_+ + Y_-,\hfill\llap{(3.1a)}\cr}$$
where
$$\displaylines{
\hspace{1em}[X_+,X_-]=0,\qquad[X_+,Y_-]=[X_-,Y_+]=0,\hfill\cr
\hspace{1em}{\rm Tr}X_+X_-={\rm Tr}\dot X_+\dot X_-=0.
\hfill\llap{(3.1b)}\cr}$$
Notice that in view of (3.1) the manifolds (2.2) take the form
\begin{equation}
\setcounter{equation}{2}%
{\rm Tr}(X_+^2 + X_-^2) = {\rm const}.
\end{equation}
We now return to (2.3).  From (2.3) and (3.1) we get
\begin{eqnarray}
&&\ddot X_+ + \beta \dot X_+ + \frac{{\rm Tr}({\dot X}_+^2+{\dot X}_-^2)}
{{\rm Tr}(X_+^2+X_-^2)}X_+ = e^{iX_+}Y_+e^{-iX_+} - Y_+,\nonumber\\
&&\ddot X_- + \beta \dot X_- + \frac{{\rm Tr}({\dot X}_+^2+{\dot X}_-^2)}
{{\rm Tr}(X_+^2+X_-^2) }X_- = e^{iX_-}Y_-e^{-iX_-} - Y_-,\\
&&X_+(0)=X_{+0},\qquad X_-(0)=X_{-0},\qquad \dot X_+(0)=
\dot X_{+0},\qquad \dot X_-(0)=\dot X_{-0}.\nonumber
\end{eqnarray}
Now consider the following realization of the generators of the
$SU(2)\times SU(2)$ group:
\begin{equation}
{\bf J}_+ =
\left(\begin{array}{c|c} \hbox{$\frac{1}{2}$}\bs & 0 \\ \hline
0 & 0
\end{array}\right),\qquad {\bf J}_- =
\left(\begin{array}{c|c} 0 & 0 \\ \hline
0 & \hbox{$\frac{1}{2}$}\bs
\end{array}\right),
\end{equation}
where $\bs=(\sigma_1,\sigma_2,\sigma_3)$
and $\sigma_i$, $i=1$, 2, 3, are the Pauli matrices.  Clearly, we can
write the general elements of the Lie algebra $X_{\pm}(t)$ and
$Y_{\pm}$ of the Lie algebra as
\begin{eqnarray}
&&X_+(t) = {\bf x}_+(t)\bd{\bf J}_+,\qquad X_-(t) =
{\bf x}_-(t)\bd{\bf J}_-,\nonumber\\
&&Y_+ = {\bf a}_+\bd{\bf J}_+,\qquad Y_- = {\bf a}_-\bd{\bf J}_-,
\end{eqnarray}
where ${\bf x}_{\pm}(t)$: ${\bf R}\to{\bf R}^3$, ${\bf a}_{\pm}$ is
a constant vector of ${\bf R}^3$ and the dot
designates the inner product.  Substituting (3.5) into (3.3) we
obtain the following nonlinear system of second-order equations:
\begin{eqnarray}
&&\ddot{\bf x}_+ + \beta\dot{\bf x}_+ + \frac{\dot{\bf
x}_+^2+\dot{\bf x}_-^2}{{\bf x}_+^2+{\bf x}_-^2}{\bf x}_+ =
(\cos|{\bf x}_+|-1){\bf a}_+ + \frac{\sin|{\bf x}_+|}{|{\bf
x}_+|}{\bf a}_+\times{\bf x}_+ + (1-\cos|{\bf
x}_+|)\frac{({\bf a}_+\bd{\bf x}_+){\bf x}_+}{{\bf x}_+^2},\nonumber\\
&&\ddot{\bf x}_- + \beta\dot{\bf x}_- + \frac{\dot{\bf
x}_+^2+\dot{\bf x}_-^2}{{\bf x}_+^2+{\bf x}_-^2}{\bf x}_- =
(\cos|{\bf x}_-|-1){\bf a}_- + \frac{\sin|{\bf x}_-|}{|{\bf
x}_-|}{\bf a}_-\times{\bf x}_- + (1-\cos|{\bf
x}_-|)\frac{({\bf a}_-\bd{\bf x}_-){\bf x}_-}{{\bf x}_-^2},\nonumber\\
&&{\bf x}_+(0) = {\bf x}_{+0},\qquad {\bf x}_-(0) = {\bf
x}_{-0},\qquad \dot{\bf x}_+(0) = \dot{\bf x}_{+0},\qquad \dot{\bf
x}_-(0) = \dot{\bf x}_{-0},
\end{eqnarray}
where ${\bf a}_{\pm}\times{\bf x}_{\pm}$ designates the vector product
of vectors ${\bf a}_{\pm}$ and ${\bf x}_{\pm}$, and $|{\bf
x}_{\pm}|=\sqrt{{\bf x}_{\pm}^2}$
is the norm of the vector ${\bf x}_{\pm}$.

Notice that the manifolds given by (3.2) and (3.5) are the
five-dimensional spheres
\begin{equation}
{\bf x}_+^2+{\bf x}_-^2 = {\rm const}.
\end{equation}

We now examine the asymptotic behaviour of the system (3.6). First
we observe that (3.6) implies
\begin{equation}
\frac{\partial}{\partial \dot{\bf x}_+}\bd\ddot{\bf x}_+ +
\frac{\partial}{\partial \dot{\bf x}_-}\bd\ddot{\bf x}_- = -6\beta
-2\frac{{\bf x}_+\bd\dot{\bf x}_++{\bf x}_-\bd\dot{\bf x}_-}{{\bf
x}_+^2+{\bf x}_-^2}.
\end{equation}
Thus as the solution to (3.6) when $\beta >0$ approaches the sphere
(3.7), the system (3.6) becomes dissipative one with exponential
contraction of a volume element.  Furthermore, it can be easily
checked that the $SU(2)\times SU(2)$ realization of (2.4) takes the
form
\begin{equation}
{\bf x}_+^2 + {\bf x}_-^2= 2({\bf x}_{+0}\bd\dot{\bf
x}_{+0}+{\bf x}_{-0}\bd\dot{\bf
x}_{-0})\hbox{$1\over\beta$}(1-e^{-\beta t}) + {\bf x}_{+0}^2+{\bf
x}_{-0}^2.
\end{equation}
Thus whenever the initial data satisfy the inequality
\begin{equation}
\hbox{$2\over\beta$}({\bf x}_{+0}\bd\dot{\bf x}_{+0}+{\bf x}_{-0}\bd\dot{\bf
x}_{-0}) + {\bf x}_{+0}^2+{\bf x}_{-0}^2 > 0,
\end{equation}
and $\beta>0$, then the solution to (3.6) approaches the
sphere $S_5$
\begin{equation}
{\bf x}_+^2 + {\bf x}_-^2= \hbox{$2\over\beta$}({\bf x}_{+0}\bd\dot{\bf
x}_{+0}+{\bf x}_{-0}\bd\dot{\bf x}_{-0}) + {\bf x}_{+0}^2+{\bf x}_{-0}^2.
\end{equation}
Evidently, the sphere $S_5$ is an invariant set for the initial conditions
such that
\begin{equation}
{\bf x}_{+0}\bd\dot{\bf x}_{+0}+{\bf x}_{-0}\bd\dot{\bf x}_{-0} = 0
\end{equation}
and arbitrary $\beta$.  The solutions corresponding to the remaining
initial data do not approach the sphere $S_5$.  Indeed, if $\beta>0$ and
\begin{equation}
\hbox{$2\over\beta$}({\bf x}_{+0}\bd\dot{\bf x}_{+0}+{\bf x}_{-0}\bd
\dot{\bf x}_{-0}) + {\bf x}_{+0}^2+{\bf x}_{-0}^2 = 0,\qquad
{\bf x}_{+0}\bd\dot{\bf x}_{+0}+{\bf x}_{-0}\bd\dot{\bf x}_{-0} \ne 0,
\end{equation}
then the solutions to (3.6) tend asymptotically to the singular
point ${\bf x}_+={\bf 0},\,{\bf x}_-={\bf 0}$.  Furthermore, if
$\beta>0$ and
\begin{equation}
\hbox{$2\over\beta$}({\bf x}_{+0}\bd\dot{\bf x}_{+0}+{\bf x}_{-0}\bd
\dot{\bf x}_{-0}) + {\bf x}_{+0}^2+{\bf x}_{-0}^2 < 0,
\end{equation}
then the singular point is approached after a finite period of time
\begin{equation}
t_* = -\frac{1}{\beta}{\rm ln}\left(1+\frac{\beta({\bf
x}_{+0}^2+{\bf x}_{-0}^2)}{2({\bf x}_{+0}\bd\dot{\bf x}_{+0}+{\bf
x}_{-0}\bd\dot{\bf x}_{-+0})}\right).
\end{equation}
Finally, for $\beta\le0$ and ${\bf x}_{+0}\bd\dot{\bf x}_{+0}+{\bf
x}_{-0}\bd\dot{\bf x}_{-0} \ne 0$, the trajectories go to infinity.

We now return to (3.6).  Notice that in view of the covariance of (3.3)
with respect to the group transformations we can set in (3.6)
without loose of generality
\begin{equation}
{\bf a}_+=(0,0,a\cos\alpha),\qquad {\bf a}_+=(0,0,a\sin\alpha).
\end{equation}
Thus we finally arrive at the system
\begin{eqnarray}
&&\ddot x_{\pm1} + \beta\dot x_{\pm1} + \frac{\dot{\bf
x}_+^2+\dot{\bf x}_-^2}{{\bf x}_+^2+{\bf x}_-^2}x_{\pm1} =
- \frac{\sin|{\bf x}_{\pm}|}{|{\bf x}_{\pm}|}a_{\pm3}x_{\pm2} +
(1-\cos|{\bf x}_{\pm}|)\frac{a_{\pm3}x_{\pm1}x_{\pm3}}
{{\bf x}_{\pm}^2},\nonumber\\
&&\ddot x_{\pm2} + \beta\dot x_{\pm2} +  \frac{\dot{\bf
x}_+^2+\dot{\bf x}_-^2}{{\bf x}_+^2+{\bf x}_-^2}x_{\pm2} =
\frac{\sin|{\bf x}_{\pm}|}{|{\bf x}_{\pm}|}a_{\pm3}x_{\pm1} +
(1-\cos|{\bf x}_{\pm}|)\frac{a_{\pm3}x_{\pm2}x_{\pm3}}
{{\bf x}_{\pm}^2},\nonumber\\
&&\ddot x_{\pm3} + \beta\dot x_{\pm3} + \frac{\dot{\bf
x}_+^2+\dot{\bf x}_-^2}{{\bf x}_+^2+{\bf x}_-^2}x_{\pm3} =
(\cos|{\bf x}_{\pm}|-1)a_{\pm3} +
(1-\cos|{\bf x}_{\pm}|)\frac{a_{\pm3}x_{\pm3}^2}
{{\bf x}_{\pm}^2},\nonumber\\
&&{\bf x}_{\pm}(0)={\bf x}_{\pm0},\qquad \dot{\bf
x}_{\pm}(0)=\dot{\bf x}_{\pm0},
\end{eqnarray}
where $a_{+3}=a\cos\alpha$ and $a_{-3}=a\sin\alpha$.  Notice
that the parameter $\alpha$ can be restricted to the interval
$[0,\pi/2]$.  Indeed, taking into account (3.16) we find that the
transformation $\alpha\to\alpha+\pi/2$ leads to
\begin{equation}
a_{+3}\to-a_{-3},\qquad a_{-3}\to a_{+3},
\end{equation}
or in view of (3.17)
\begin{equation}
{\bf x}_-\to {\bf x}_+,\qquad x_{+1}\to-x_{-1},\qquad x_{+2}\to
x_{-2},\qquad x_{+3}\to-x_{-3}.
\end{equation}
Further, we observe that for $\alpha=\pi/4$ the system (3.17) is
symmetric in ${\bf x}_+$ and ${\bf x}_-$ variables.  For an easy
illustration of this observation see Fig.\ 2.  It should also be
noted that by virtue of the following transformation law of the
equation (2.3) under the scaling $t\to\lambda t$:
\begin{equation}
\ddot X + \lambda\beta \dot X + \frac{{\rm Tr}{\dot X}^2}{{\rm
Tr}X^2 }X = e^{iX}\lambda^2Ye^{-iX} - \lambda^2Y,
\end{equation}
we have actually two bifurcation parameters: $\beta^2/a$, where
$a\ne0$, and $\alpha$.

In order to discuss the symmetries of (3.17) consider the original
abstract equation (2.3).  Evidently, (2.3) is invariant under the
transformations referring to the stability group of $Y$, that is
transformations leaving $Y$ unchanged.  Hence, making use of
(3.3) and (3.5) we find that solutions to (3.17) have the form invariant under
rotations about $x_{+3}$ and $x_{-3}$ axes such that
\begin{eqnarray}
&&x'_{\pm1} = \cos\varphi_{\pm} x_{\pm1} + \sin\varphi_{\pm} x_{\pm2},\nonumber\\
&&x'_{\pm2} = -\sin\varphi_{\pm} x_{\pm1} + \cos\varphi_{\pm} x_{\pm2},\nonumber\\
&&x'_{\pm3} = x_{\pm3},\qquad \varphi_{\pm}\in[0,2\pi).
\end{eqnarray}
This suggests that whenever the initial condition ${\bf x}_{+0}$, ${\bf
x}_{-0}$, $\dot{\bf x}_{+0}$ and $\dot{\bf x}_{-0}$ corresponds to
an attractor $A$ then the basin of attraction of $A$ contains the
points ${\bf x}'_{+0}$, ${\bf x}'_{-0}$, $\dot{\bf x}'_{+0}$ and
$\dot{\bf x}'_{-0}$ of the form
\begin{subequations}
\begin{eqnarray}
&&x'_{\pm10} = \cos\varphi_{\pm} x_{\pm10} + \sin\varphi_{\pm} x_{\pm20},\nonumber\\
&&x'_{\pm20} = -\sin\varphi_{\pm} x_{\pm10} + \cos\varphi_{\pm} x_{\pm20},\nonumber\\
&&x'_{\pm30} = x_{\pm30},\\
&&\dot x'_{\pm10} = \cos\varphi_{\pm}\dot x_{\pm10} + \sin\varphi_{\pm}\dot x_{\pm20},\nonumber\\
&&\dot x'_{\pm20} = -\sin\varphi_{\pm}\dot x_{\pm10} + \cos\varphi_{\pm}\dot x_{\pm20},\nonumber\\
&&\dot x'_{\pm30} = \dot x_{\pm30},\qquad \varphi_{\pm}\in[0,2\pi).
\end{eqnarray}
\end{subequations}
Thus the basin of attraction contains two circles given by (3.22a)
such that
\begin{equation}
{\bf x'}_{+0}^2={\bf x}_{+0}^2,\qquad {\bf x'}_{-0}^2={\bf x}_{-0}^2,
\end{equation}
and two vector fields (3.22b) obtained from $\dot{\bf x}_{\pm0}$ by
rotating this about $x_{+3}$ and $x_{-3}$ axis, respectively, by the
angle referring to the position of the corresponding point of the
circle.

In figures 1 and 2 we show examples of strange
attractors from numerical integration of the system (3.17).  As
expected these attractors are symmetric under rotations about
$x_{+3}$ and $x_{-3}$ axes.\\[\baselineskip]
\centerline{\fbox{Fig.\ 1, 2}}\\[\baselineskip]
It follows from the computer simulations illustrated in Fig.\ 3
that in the parameter space of the system (3.17) in a neighbourhood
of the attractor from Fig.\ 1 there exists a quasiperiodic
trajectory.\\[\baselineskip]
\centerline{\fbox{Fig.\ 3}}\\[\baselineskip]
A look at Fig.\ 3 is enough to conclude that in the case with the attractor
from Fig.\ 1 we deal with the quasiperiodicity to chaos transition like in
the Ruelle-Takens-Newhouse scenario [2,3].  In the case of the attractor
from Fig.\ 2 we have most probably the new scenario.\\[\baselineskip]
\centerline{\fbox{Fig.\ 4--6}}\\[\baselineskip]
Namely, it turns out that there exist in the parameter space of the
system (3.17) two nearby quasiperiodic orbits shown in Fig.\ 4.
The computer simulations suggest that these quasiperiodic orbits are
separated by an infinitesimal perturbation of the bifurcation
parameter $\beta$.  As a consequence of the infinitesimal nature of
the perturbation the irregular transitions occur between the chaotic
attractors arising from the perturbation of the quasiperiodic orbits
from Fig.\ 4a and Fig.\ 4b, respectively.  These attractors are
shown separately in Fig.\ 6.  We remark that the attractors have
the same form as those shown in Fig.\ 5 arising in the transient
chaos before reaching the quasiperiodic state.  The authors did not find such
scenario of transitions between chaotic attractors in the
literature.  As the bifurcation parameter decreases and approaches
the value corresponding to the attractor from Fig.\ 2 the frequency
of transitions between the two attractors increases.  On the other
hand, the decay of the bifurcation parameter leads to the
deformation of the attractors.  As a result of these two combined
processes we arrive at the attractor illustrated in Fig.\ 2.  In
this sense that attractor arise from the perturbation of two
quasiperiodic states.

\vspace{1.5\baselineskip}
\centerline{\bf 4.\ CONCLUSION}

\vspace{\baselineskip}
In the present work we have studied the particular $SU(2)\times
SU(2)$ realization of the abstract Newton-like equation having
attractors coinciding with the submanifolds fixed by the
second-order Casimir operator referring to a given Lie group.  The
resulting second-order system in ${\bf R}^6$ has been shown to
exhibit chaotic behaviour.  The concrete examples of the chaotic
attractors with the $SU(2)\times SU(2)$ symmetry have been provided.
One of them (see Fig.\ 2 and discussion below) is related to the
most probably new scenario of transitions between chaotic attractors
and effectively forming one by perturbation of two quasiperiodic
states.  We point out that the case with the $SU(2)\times SU(2)$
realization of the abstract Newton-like equation discussed herein
seems to be the simplest one leading to chaotic dynamics.  As far
as we are aware the investigated system (3.17) is the first example
of the chaotic dynamical system on the manifold determined by the
continuous Lie-group structure.  We remark that the importance of
such an example was indicated in [4--6].  Furthermore, the knowledge
of chaotic attractors with fixed symmetry and known the invariant
measure like those introduced in this work would be useful for
testing methods of detecting symmetry of chaotic attractors such as
for example the method of detectives [7].  Whenever the theory of
groups appears to lead to some insight into the nature of chaos
then it seems that the observations introduced herein would be an
important point of departure in solving numerous problems.

\vspace{1.5\baselineskip}
\centerline{\bf ACKNOWLEDGEMENT}

\vspace{\baselineskip}
This work was supported by UL grant 488.
\newpage
\centerline{\bf REFERENCES}

\vspace{\baselineskip}
\noindent 1.\ K. Kowalski and J. Rembieli\'nski, {\em Physica D\/} {\bf 99}, 237 (1996).\\[.1cm]
2.\ S. Newhouse, D. Ruelle and F. Takens, {\em Commun. Math. Phys.\/} {\bf 64}, 35 (1978).\\[.1cm]
3.\ D. Ruelle and F. Takens, {\em Commun. Math. Phys.\/} {\bf 20}, 167 (1971).\\[.1cm]
4.\ J.W. Swift and E. Barany, {\em Eur. J. Mech. B/Fluids\/} {\bf 10}, 99 (1991).\\[.1cm]
5.\ J. Guckenheimer and P. Worfolk, {\em Nonlinearity\/} {\bf 5}, 1211 (1992).\\[.1cm]
6.\ M. Field and J.W. Swift, {\em Nonlinearity\/} {\bf 7}, 385 (1994).\\[.1cm]
7.\ E. Barany, M. Dellnitz and M. Golubitsky, {\em Physica D\/} {\bf 67}, 66 (1993).
\newpage
\noindent {\large Legends to figures}\\[2\baselineskip]
\noindent Fig.\ 1.  The system (3.17) with $\beta=0.1$, $a=0.5$,
$\alpha=0.2$, ${\bf x}_{+0}=(1,1,1.5)$, ${\bf x}_{-0}=(1,1,1.5)$,
$\dot{\bf x}_{+0}=(-0.1,0.1,0.1)$ and $\dot{\bf
x}_{-0}=(0.1,0.1,0.1)$.  Left: the projection of the Poincar\'e
section of the attractor on the $(x_{+1},x_{+3})$ plane.  This
section and the following ones in the case with the projection on
the $(x_{+1},x_{+3})$ plane are defined by the hyperplane
$n(x-x^{(0)})=0$, where $n$ is the normal vector and $x^{(0)}$ is a
point of the hyperplane, $n,\,x^{(0)},\,x\in{\bf R}^{12}$,
$n=(0,1,0,\ldots,0)$ and $x^{(0)}=0$.  Right: the projection of the Poincar\'e
section of the attractor on the $(x_{-1},x_{-3})$ plane.  This
section and the following ones in the case with the projection on
the $(x_{-1},x_{-3})$ plane are defined by the hyperplane
$n(x-x^{(0)})=0$, where $n=(0,0,0,0,1,0,\ldots,0)$ and $x^{(0)}=0$.
The Lyapunov exponents are: $\lambda_1=0.02$,
$\lambda_2=\lambda_3=\lambda_4=\lambda_5=\pm0.00$, $\lambda_6=-0.01$,
$\lambda_7=-0.06$, $\lambda_8=-0.08$, $\lambda_9=-0.09$,
$\lambda_{10}=-0.10$, $\lambda_{11}=-0.12$ and $\lambda_{12}=-0.16$.
The Lyapunov dimension is 6.1.\\[\baselineskip]
\noindent Fig.\ 2.  The system (3.17) with $\beta=0.1$, $a=1$,
$\alpha=\pi/4$, ${\bf x}_{+0}=(2,2,2)$, ${\bf x}_{-0}=(2,2,2)$,
$\dot{\bf x}_{+0}=(0.1,0.1,-1)$ and $\dot{\bf x}_{-0}=(1,1,1)$.
Left: the projection of the Poincar\'e
section of the attractor on the $(x_{+1},x_{+3})$ plane.    Right:
the projection of the Poincar\'e
section of the attractor on the $(x_{-1},x_{-3})$ plane.
The Lyapunov exponents are: $\lambda_1=0.06$, $\lambda_2=0.01$,
$\lambda_3=\lambda_4=\lambda_5=\lambda_6=\pm0.00$, $\lambda_7=-0.04$,
$\lambda_8=-0.07$, $\lambda_9=-0.10$, $\lambda_{10}=-0.11$,
$\lambda_{11}=-0.14$ and $\lambda_{12}=-0.21$.  The Lyapunov dimension
is 7.4.\\[\baselineskip]
\noindent Fig.\ 3.  The system (3.17) with $\beta=0.117$. The
remaining parameters and the initial condition are the same as in
Fig.\ 1.  Left: the projection of the Poincar\'e
section of the quasiperiodic attractor on the $(x_{+1},x_{+3})$
plane.    Right: the projection of the Poincar\'e
section of the quasiperiodic attractor on the $(x_{-1},x_{-3})$
plane.\\[\baselineskip]
\noindent Fig.\ 4.  The system (3.17) with a) $\beta=0.1102766$ and
b) $\beta=0.1102767$. The
remaining parameters and the initial condition are the same as in
Fig.\ 2.  The projection of the Poincar\'e
section of the quasiperiodic attractors on the $(x_{+1},x_{+3})$
plane.\\[\baselineskip]
\noindent Fig.\ 5.  The transient chaos in the system (3.17) with a)
$\beta=0.1102766$ and b) $\beta=0.1102767$ before reaching the
quasiperiodic states from Fig.\ 4. The
remaining parameters and the initial condition are the same as in
Fig.\ 2.  The projection of the Poincar\'e
section of the attractors on the $(x_{+1},x_{+3})$
plane.\\[\baselineskip]
\noindent Fig.\ 6.  The system (3.17) with  $\beta=0.1099$.  The
remaining parameters and the initial condition are the same as in
Fig.\ 2.  For such data the irregular transitions appear between the
attractor from the left and the attractor from the right.  In both
figures the projection is shown of the Poincar\'e
section of the attractor on the $(x_{+1},x_{+3})$
plane.
\end{document}